\documentclass[
groupedaddress,floatfix,
 amsmath,amssymb,twocolumn,
 aps,
prl]{revtex4-2}

\usepackage{graphicx}
\usepackage{dcolumn}
\usepackage{bm}
\usepackage{hyperref}
\usepackage{multirow}
\usepackage[table]{xcolor}

\usepackage[skip=3.5pt]{parskip}
\usepackage{hyperref}
\usepackage{color}
\usepackage{subfigure}
\usepackage{flexisym}
\usepackage{xcolor}

\begin{document}
\title{Measurement of the motional heating of a levitated nanoparticle by thermal light}


\author{A. T. M. Anishur Rahman}
 \altaffiliation{Department of Physics \& Astronomy, University College London, London, UK}
 \altaffiliation{Present address: Department of Physics, University of Warwick, Coventry, UK}
 \email{a.rahman@ucl.ac.uk}
\author{P. F. Barker}%
 \altaffiliation{Department of Physics \& Astronomy, University College London, London, UK}
\email{p.barker@ucl.ac.uk}


\begin{abstract}

We report on measurements of photon induced heating of silica nanospheres levitated in vacuum by a thermal light source formed by a superluminescent diode. Heating of the nanospheres motion along the three trap axes was measured as a function of gas pressure and for two particle sizes. Heating rates were also compared with the much lower reheating of the same sphere when levitated by a laser. We find the measured trap heating rates are dominated by the much larger heating rates expected from the recoil of thermal photons.
\end{abstract}
\maketitle



 Optomechanical interactions are routinely used to cool and control the motion of objects that range from the kilogram mass scale down to the atomic scale. Important applications include gravitational wave detection \cite{AbbottPRL2016}, but also the trapping of atoms in focused laser beams\cite{GRIMM200095}. These interactions, are however eventually limited by the discrete nature of the light. For example, fluctuations in photon number lead to measurement noise or imprecision, while the backaction of the light on the object leads to radiation pressure shot noise or recoil heating\cite{kurn,Jain2016,PurdySciene2013}. Stochastic heating from the photon recoil eventually leads to trap loss if not compensated by some form of cooling. In far-off resonant traps for cold atoms formed by focused laser beams, recoil limited heating rates below the millikelvin per second range have been achieved \cite{MillerPRA1993} and trap lifetimes of the atomic ensembles up to hundreds of seconds range have been demonstrated by minimizing the scattered light \cite{GRIMM200095,OHaraPRL1999}. We note that these however typically limited by classivcal intensity noise the laser pointing instability. The optomechanical effects of the photon statistics from different types of sources such as thermal or squeezed light sources on levitated systems has not yet been explored, although squeezed light sources are now used to to reduce measurement noise in laser inteferometric gravitational wave detection \cite{AasiJ2013Esot}.
 
 
 Recently, we have demonstrated the trapping of nanoparticles with intense thermal light from a superluminescent diode (SLD)\cite{RahmanOptica2020}. The broadband nature of this source allows spectral shaping of the output profile which can in principle be used to create arbitrary optical potentials. SLD's are also widely used in optical coherence tomography \cite{HuangScience1991}, ghost imaging \cite{GattiPRL2004}, fiber optic gyroscopes \cite{BohmK1981} and other emerging applications where intense, low coherence light is required \cite{Kafar_2019,Hartmann2017,BoitierNatPhys2009}. The low temporal coherence time and the Bose-Einstein distributed thermal photon statistics of such light sources can also be described by a blackbody source with a chemical potential \cite{Wurfel_1982,Henry1986}. These sources have significantly increased quantum noise when compared to a laser \cite{Hartmann_2015}. Moreover, as this light has a thermal distribution the motional temperature of any levitated object should equilibrate to the temperature of the light source which can be controlled and used potentially to passively determine the temperature of the trapped particle. This is the type of damping originally envisioned by Einstein when understanding the nature of Blackbody (BB) radiation \cite{Einstein1909}. Unlike a BB source, an SLD has a high spatial coherence with a well-defined polarization \cite{Hartmann2017,GattiPRL2004} allowing it to be tightly focused and used for optical trapping and levitation as has recently demonstrated.\cite{RahmanOptica2020}.

\begin{figure*}
    \centering
    \includegraphics[width=17.5cm]{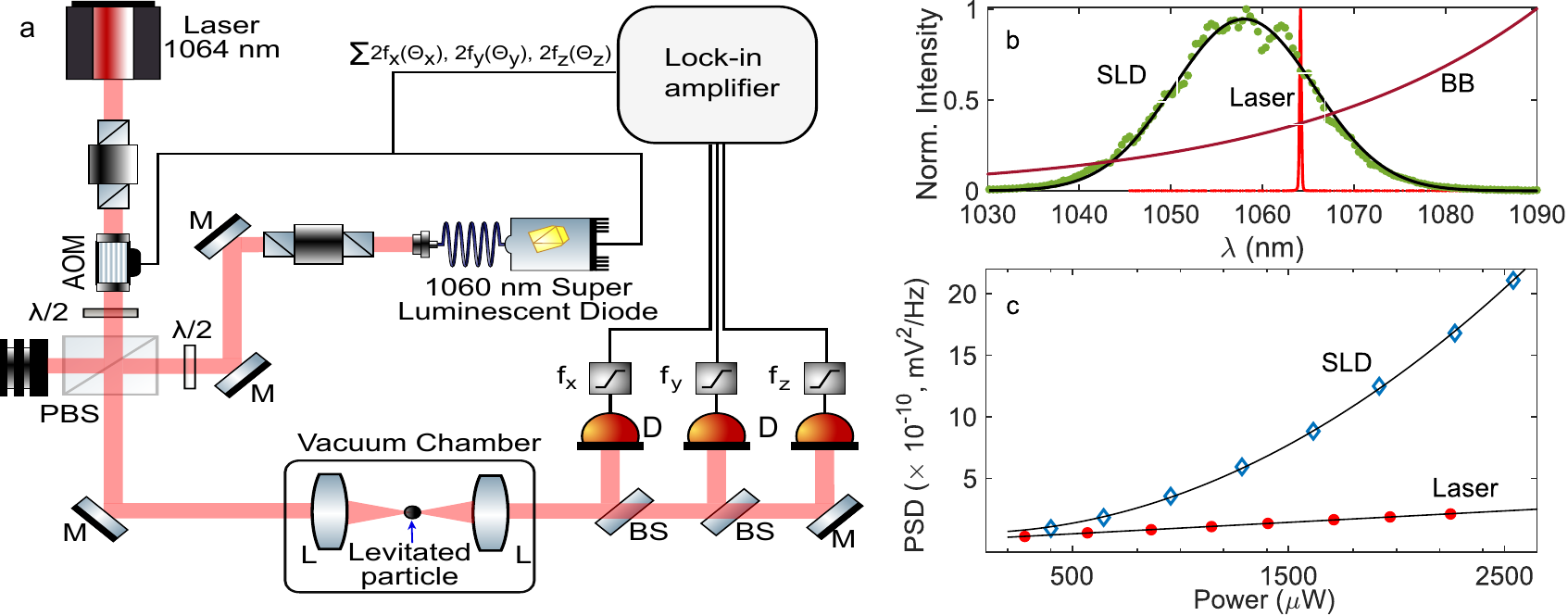}
    \caption{Experimental details- a) A schematic of our experimental setup in which a particle can be levitated and parametrically cooled using a laser beam or the light from a superluminescent diode. Different components are - D balanced photodiode, M mirror, BS beam splitter, L lens, PBS polarizing beam splitter, $\lambda/2$ half wave plate. For parametric feedback signals from photodiodes are sent to lock-in amplifiers which after appropriate processing generate signals at double the trap frequency and modulate the intensity of either the laser beam or SLD beam. b) Normalized spectral intensity profiles of the superluminescent diode, a single longitudinal mode laser and an ideal blackbody source (BB). The spectral profile of the SLD has been fitted with a Gaussian (black solid line). c) Shot noise as a function of input power to the balanced photodiode for the laser as well as the SLD beam. Each data point in this graph represents spectral density at $200~$kHz for different input power to a balance photodiode. Solid black lines represent fits. For the case of laser, the fit represent a line of the form $a+bP$ while for the SLD case the line is of the form $a+bP^2$, where $a$ and $b$ constants.}
    \label{fig0}
\end{figure*}
We report on the measurement of the heating of a levitated nanomechanical oscillator in high vacuum when levitated using a superluminescent diode. In particular, we confirm the enhanced heating and damping of the oscillator expected from such a source when compared with the same oscillator levitated by a laser. We also describe the measurement of the direction dependent heating along all three axes of the levitated nanoparticle which is consistent with heating dominated by the scattering of thermal photons.


To measure the heating of the levitated particle by thermal light we create an optical tweezer by tightly focusing the collimated beam of each source using a single high numerical aperture lens (NA=0.77). Figure \ref{fig0}a shows the optical arrangement of our tweezer inside a vacuum chamber. The laser operates at a wavelength of $1064~$nm, while the superluminescent diode is centered around $1060$ nm. We trap a silica nanoparticle using one of the beams and then transfer it to the trap created by the other beam when required \cite{RahmanOptica2020}. This transfer process is carried out at a pressure of $\approx 5~$mBar by reducing the power of one beam while increasing in the other. The spectral intensity profiles of our SLD and the laser are shown in Fig. \ref{fig0}b. 

The light from the SLD is the result of amplified spontaneous emission from a biased p-n junction \cite{Hartmann2017,Blazek2012}. To avoid optical feedback, and hence lasing, the facets of the SLD are at an angle and anti-reflection coated. The spectral intensity profile of spontaneous emission from a p-n junction \cite{Wurfel_1982} has the form $\frac{G(\omega)\hbar\omega^3}{\pi^2c^2}\frac{1}{\exp{[(\hbar\omega-\mu_c)/k_BT]}-1}$, where $\omega$ and $c$ are the angular frequency and the speed of light in vacuum, $k_B$ is the Boltzmann constant and $T$ is the bulk temperature of the light source. The factor $\frac{1}{\exp{[(\hbar\omega-\mu_c)/k_BT]}-1}$ is the average photon number $\bar{n}$ in a mode, while $\frac{\hbar\omega^3}{\pi^2c^2}$ represents the degeneracy of that mode. The factor $G(\omega)$ takes into account a non-uniform spectral gain with frequency and modification to the degeneracy factor. In our SLD $G(\omega)$ has approximately a Gaussian profile. The chemical potential $\mu_c$ determines the energy stored in the electron-hole gas in the biased p-n junction \cite{Wurfel_1982}. Equivalently, it is also the energy gap between the valance and conduction bands in a semiconductor \cite{Henry1986}. The energy of the photons emitted by a SLD is greater than $\mu_c$, where $\hbar\omega > \mu_c\gg k_BT$. The photon statistics of a SLD are that of a thermal source \cite{LiZQ2010,Henry1986,Hartmann2017,Blazek2012,Hartmann_2015} where the variance of the photon number is given by $\Delta n^2=\bar{n}+\bar{n}^2$. This is in contrast to the variance of the photon number of a single mode laser e.g. $\Delta n^2 =\bar{n}$.




The spectral density $S_{vv}$ of a light source at a given power $P$ ($\propto \bar{n}$) is proportional to $\Delta n^2$ \cite{QuantumOpticsFox,Blazek2012}. Figure \ref{fig0}c shows the noise power spectral density (PSD) measured using a balanced photodiode of our laser and SLD at $200~$kHz as a function of input power. As expected, the spectral density of the laser increases linearly with input power as shown by the fit ($a+bP$) given by the black solid line, whereas the SLD light has the expected quadratic dependence (black line of the form $a+bP^2$, where $a$ and $b$ are constants). Importantly, despite its spectral profile being modified from that of an ideal blackbody source by the gain profile, the  SLD with chemical potential reproduces the power spectral density of a thermally distributed Bose-Einstein photon source.

After trapping a silica nanoparticle at atmospheric pressure, we reduce the chamber pressure and cool its center-of-mass motion using parametric feedback cooling \cite{Gieseler2012,RahmanOptica2020}. The motional dynamics of the particle along all three axes, including the effect of feedback cooling and the photon recoil, can be described by the Langevin equation,
\begin{eqnarray}
M\frac{d^2q}{dt^2}+M(\gamma_g+\gamma_{ph}^q+\gamma_{fb})\frac{dq}{dt}+M\omega_q^2, q&=&f(t),
\end{eqnarray}

where $q$ is the displacement along the $x$, $y$ or $z$ axis, and $f(t)$ is a zero mean Gaussian random process with a variance $2Mk_B(\gamma_g T_g+\gamma_{ph}^q T_{ph}^q)$. The particle has mass $M$ and oscillation frequency $\omega_q$ along the axis $q$. The damping rate due to the thermal bath of the background gas at a temperature $T_g$ is given by $\gamma_g$. In contrast, $\gamma_{ph}^q$ and $T_{ph}^q$ are the damping and the temperature of the photon bath \cite{RahmanPRL2021,NovotnyPRA2017}. Due to the directional nature of photon scattering, $\gamma_{ph}^q$ is axis dependent \cite{NovotnyPRA2017,SebersonPRA2020} and, for a spherical particle, depends on the polarization of the trapping light. The damping exerted by the feedback is given by $\gamma_{fb}$.

To estimate the recoil heating due to the thermal SLD source we calculate the gain in energy of the levitated particle through the fluctuation of the photon number given by \cite{RahmanPRL2021}
\small
\begin{eqnarray}
\nonumber
\frac{dE}{dt}|_{ph}^q&=& \frac{A_{pn}G}{2MA_w\Omega_{mx}}\int_0^\pi\int_0^{2\pi}\int_0^{\theta_{mx}}\int_0^{2\pi}\int_{\omega_c}^{\infty}P_r\sigma_{s}\frac{\omega^2~\Delta n^2}{\pi^2c^2}\\
\nonumber
&&\times \hbar^2 k^2\Biggl(\mathbf{\Theta}_{i}+\mathbf{\Theta}_{s}\Bigg)^2d\omega d\Omega_{i}d\Omega_s \\
&\approx& \frac{1}{(1-\exp[{(\mu_c-\hbar\omega_c)/k_BT}])}~\gamma_{ph}^q k_BT,
\label{eqn7}
\end{eqnarray}
\normalsize
where $\mathbf{\Theta_i}=[\sin\theta_i\cos\phi_i~ \sin\theta_i\sin\phi_i~\cos\theta_i]$ and $\mathbf{\Theta_s}=[\sin\theta_s\cos\phi_s~ \sin\theta_s\sin\phi_s~\cos\theta_s]$ are the projections of the unit vectors parallel to the incident and the scattered photons on the three translational axes. $\gamma_{ph}^q=\Lambda^q~ \frac{\sigma_cI}{Mc^2}\frac{\hbar \omega_c}{k_BT}$ with $\Lambda^q=[0.12~0.22~0.65]$, $\theta_{mx}$ is the angle between the wavevector of the incident photon and the $z-$axis and we take $\theta_{mx}=0.43$ rad. In order to obtain an analytical solution to compare with the recoil heating from a laser we set $G(\omega)$ to be constant. Our numerical solutions obtained using the actual profile of our SLD show that calculations performed using Eq. (\ref{eqn7}) are only $15\%$ higher. The damping/heating rate $\gamma_{ph}^q$ has the same form of a laser \cite{NovotnyPRA2017}, but for the SLD it is enhanced by the factor $\frac{\hbar\omega_c}{k_BT}$. The maximum solid angle $\Omega_{mx}$ that an incident photon makes with the particle is $2\pi(1-\cos\theta_{mx})$, $P_r=\frac{3}{8\pi}(\cos^{2}\theta_s\cos^2\phi_s+\sin^2\phi_s)$ is the spatial distribution of scattered photons \cite{SebersonPRA2020}, $d\Omega_i=\sin\theta_i d\phi_i d\theta_i$ and $d\Omega_s=\sin\theta_s d\phi_s d\theta_s$. The frequency of the lowest energy photon in the SLD emission spectrum is given by $\omega_c$ while $\sigma_c$ is the scattering cross section of the particle at $\omega_c$. The intensity of the thermal light at the focus is  $I=\frac{A_{pn}G\exp{[(\mu_c-\hbar\omega_c)/k_BT]}k_BT\omega_c^3}{A_w\pi^2c^2}$, where $A_{pn}$ is the surface area of the light emitting p-n junction and $A_w$ is the area of the cross section of the trapping beam at the focal spot.



\begin{figure}
    \centering
    \includegraphics[width=8.5cm]{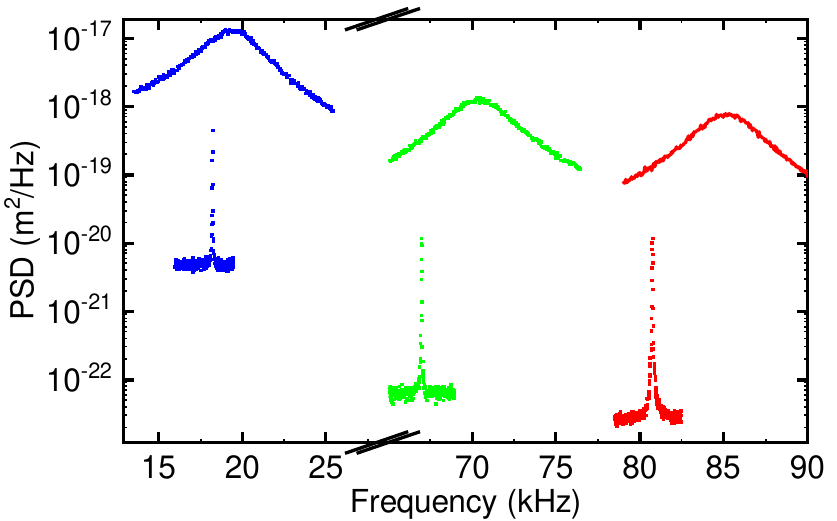}
    \caption{Power spectral densities (PSD) of $55\pm 10$ radius silica nanosphere levitated using a SLD. The trapping power at the focus was $130~$mW. Top graphs show PSDs at $5~$mbar while the bottoms graphs are the PSDs at $5\times 10^{-8}~$mbar.}
    \label{fig1}
\end{figure}

The typical sub-millihertz value of $\gamma_{ph}^q$ means that it is challenging to measure radiation damping or heating from the linewidth of the power spectral density of displacement \cite{PontinPRR2020}. This is measured by cooling the particle in high vacuum ($\gamma_g < \gamma_{ph}^q$) and subsequently allowing it to evolve towards the equilibrium while measuring the rate at which it approaches equilibrium. The evolution towards equilibrium from the initial cooled state can be described using the Fokker-Planck equation \cite{Jain2016} e.g. $T(t)_{cm}^q=T_{\infty}^q+(T_i^q-T_{\infty}^q)e^{-\gamma_q t}$, where $T_i^q$ and $T_{\infty}^q$ are the initial steady state and the equilibrium temperatures of the particle. $T_{\infty}^q$ is given by $(T_{ph}^q\gamma_{ph}^q+T_{g}\gamma_{g})/\gamma_q$ with $\gamma_q=\gamma_g+\gamma_{ph}^q$. The thermalization time ($2\pi/\gamma_q$) is typically very long and as levitated particles are extremely sensitive to external noises a prolong measurement time can reduce fidelity of measurements. We instead perform the experiment within the linear regime of the evolution process \cite{Jain2016} i.e. $t\ll 2\pi/\gamma_q$, where we have $T(t)_{cm}^q=T_i^q+(T_{\infty}^q-T_i^q)\gamma_q t$.

\begin{figure}
    \centering
    \includegraphics[width=8.5cm]{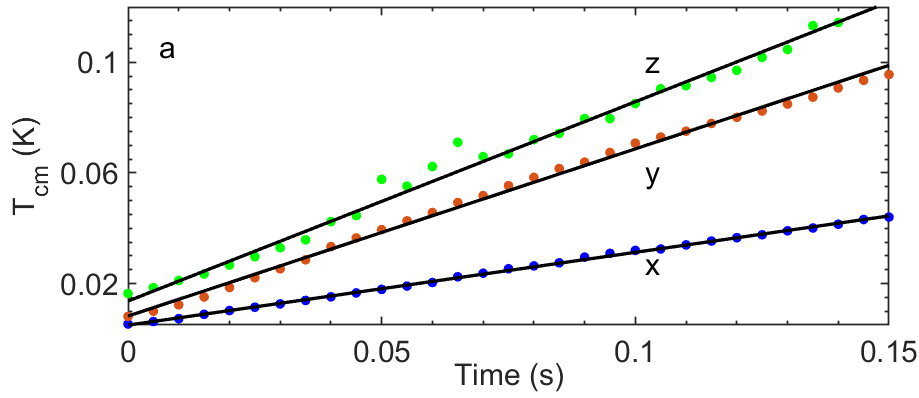}
    \centering
    \includegraphics[width=8.4cm]{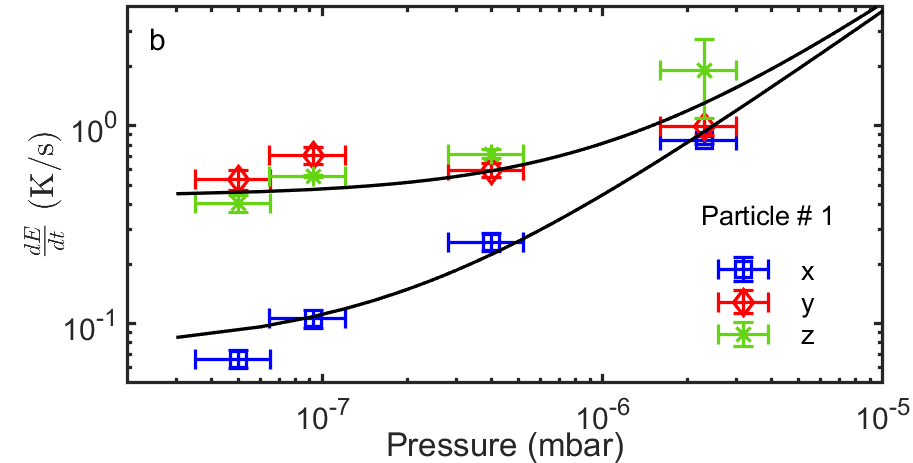}
    \caption{Effect of photon recoil- a) The evolution of the center-of-mass temperature of a $r=55 \pm 12~$nm radius silica nanoparticle at $4\times 10^{-7}~$mbar after parametric feedback is switched off at time zero. Each data point is the average of $600$ time traces. The $x-$axis represents the direction parallel to the light field polarization direction while the $y-$axis is orthogonal to the light field polarization direction. The $z-$axis is the propagation direction of the trapping light beam. Solid lines represent lines of the form $a_0+a_1t$, where $t$ denotes time, $a_0$ is the offset and $a_1$ is the reheating rate or the rate at which energy is added to the particle. b) Reheating rates as functions of gas pressure along all three axes. Solid lines represent $a_{ph}^q+a_2P_g$, where $a_{ph}^q$ is the heating rate due to the recoil of photons along the axis $q$, $P_g$ is the gas pressure inside the vacuum chamber and $a_2$ is a proportionality constant.}
    \label{fig2}
\end{figure}

To do this, we parametrically \cite{Gieseler2012,Vovrosh2017,RahmanOptica2020} cool the CM motion of a $55\pm 16~$nm radius silica sphere to sub-kelvin temperatures. The particle size was obtained from the linewidth of the position power spectral density at $5~$mBar. The uncertainty in the particle size arises from the $30\%$ uncertainty in our measurement of pressure. Figure \ref{fig1} shows power spectral densities of this particle at $5~$mBar (top graphs) and where we start reheating measurements at a pressure of $5\times 10^{-8}~$mbar (bottom graphs). At the lowest pressures the center-of-mass temperatures along the $x$, $y$ and $z$ axes were $55~$mK, $22~$mK and $45~$mK, respectively. These were found using the area under the position power spectral density graphs calibrated to the higher pressure data. 
Once the lowest temperature was reached, the parametric feedback is switched off and the particle's dynamics are monitored as a function of time as it heats up \cite{Jain2016}. After a short time, typically 150 ms, feedback is reactivated and the particle is again cooled to the well-defined minimum temperature. Parametric feedback cooling is again switched off allowing the particle to reheat. This process is repeated $600$ times and the averaged results are shown in Fig. \ref{fig2}a at a pressure of $4\times 10^{-7}~$mbar. We determine the energy, normalized by $k_B$, of the particle as a function of time from particle position using the relation $T_{cm}(t)^q=M\omega_q^2\langle q(t)^2\rangle/k_B$. The solid lines shown in figure 3a are fits of the form $a_0^q+a_1^qt$, where $a_0^q$ is the initial CM temperature of the particle along each axis $q$ and where $a_1^q$ is the reheating rate. The derived $a_1^q$ is the sum of the reheating rates due to collisions with gas molecules and optical/photonic sources of heating. The rate of increase of $T_{cm}(t)$ along the $y-$axis (orthogonal to the light field polarization) and the $z-$axis (trapping field propagation direction) are approximately equal and are significantly higher than that along the $x-$axis (light field polarization direction). Fig. \ref{fig2}b, are plots of reheating rates of the particle along all three axes as a function of residual gas pressure $P_g$. As the gas pressure increases, the directional dependence of the reheating rates diminishes as expected and eventually becomes negligible at a pressure of $2\times 10^{-6}~$mBar as the heating rate due to the gas molecules dominates that due to the photons. We fit a function $a_{ph}^q+a_2P_g$ to the reheating data as a function of gas pressure $P_g$, where $a_{ph}^q$ is the reheating rate due to the recoil of photons given by Eq. (\ref{eqn7}) and $a_2$ is the pressure dependent heating rate due to the gas molecules. We fit one line for the $x~$direction and one for $y$ and $z$ directions which have similar values. The photonic heating rates given by the fit are $a_{ph}^x= 0.08\pm 0.01~$K/s along the $x-$axis and $a_{ph}^{y,z}=0.45\pm 0.07~$K/s along the $y$ and $z$ axes. From the fit we retrieve a reheating rate ($a_2P_g$) of $0.018\pm 0.006~$K/s due to the gas molecules at $P_g=5\times 10^{-8}~$mBar. This is consistent with that calculated for a particle of radius $55$ nm. We calculate the photonic heating rates only from recoil of thermal photons using our model of Eq (\ref{eqn7}). These values are $0.07\pm 0.07~$K/s, $0.21\pm 0.15~$K/s $0.48\pm 0.50~$K/s along the $x,~y~ \&~ z$ axes respectively. The values for the $x$ and $z$ axes agree within the uncertainty of each measurement, but the heating rate in the $y$ direction is higher than expected from just recoil of thermal photons using our simple model. The relatively good agreement between the calculated and experimentally derived values for the $x$ and $z$ axes indicates that the heating rates at low pressures in these axes are dominated by recoil of thermal photons and not additional heating sources such as parametric heating. Moreover, the reheating rate due to the relative intensity noise which can lead to parametric heating was measured for the SLD to be $-115~$dB/Hz at $100$ kHz. The parametric heating from this noise was calculated \cite{Jain2016} to be $5$ times less than that due to the thermal photon. To calculate the theoretical values we have used a trapping power of $130~$mW, $2\pi\hbar c/\mu_c=1115~$nm and $2\pi c/\omega_c=1090~$nm.



The axial variation of the heating rate is also a signature of recoil heating. This results from the directional nature of the incoming field and the outgoing dipolar radiation pattern of the scattered field. Although the recoil heating from a thermal source is significantly larger than from a laser due to the much larger fluctuation in photon number, the ratio of these heating rates (normalized here by the $x$-axis value) are only dependent on the spatial distribution of the incoming and outgoing scattered field along the $x$, $y$ and $z$ axis. In the Rayleigh limit, where the particle radius is much less than the wavelength, the ratio of these heating rates has been calculated for the case of laser reheating to be [1,2,7] for a plane wave \cite{SebersonPRA2020}. For a laser operating at a wavelength of $1064~$nm and focused by a $0.77$ NA aperture lens these ratios become $[1.0,~1.7,~4.6]$. We consider only the Rayleigh regime in our calculation and obtain the same ratios as calculated for the laser for plane wave illumination and $[1.0,~1.81,~5.4]$ for the focused beam ($0.77$ NA) using our simple model. We note that this model does not take into account the known chromatic aberration at the focus due to the broadband nature of the SLD or the wave nature of light \cite{RahmanOptica2020}. The measured ratio of the heating rates are $[1,~5.6,~ 5.6]$ which, within the uncertainty of our measurements, are consistent with the predictions in the the $x$ and $z$ directions but are not in agreement for the $y-$axis which indicates that there is again likely to be excess heating on this axis. 

\begin{figure}
    \centering
    \includegraphics[width=8.6cm]{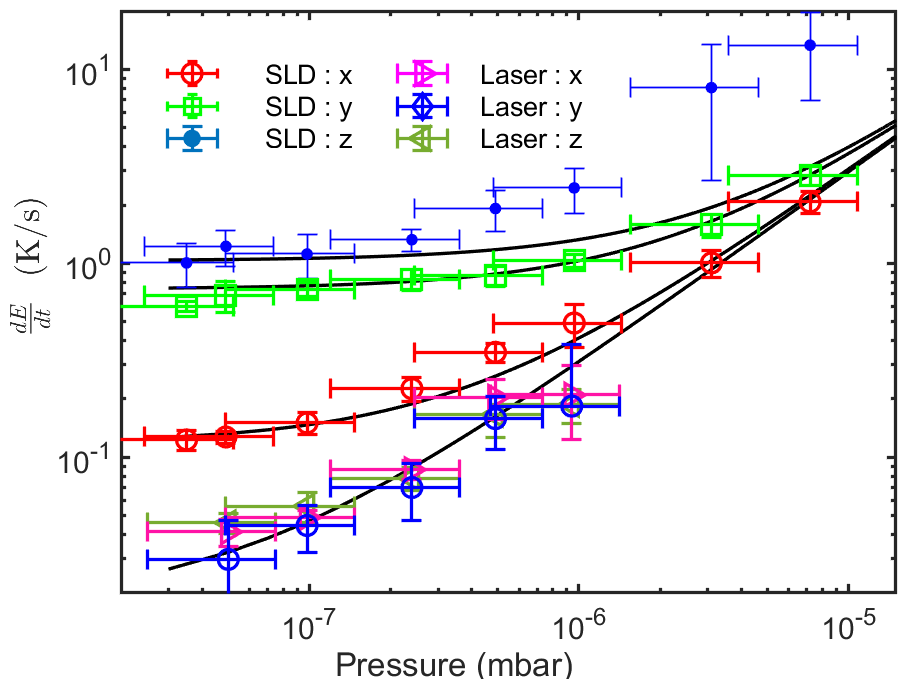}
    \caption{Reheating rates of a $r=70\pm 12~$nm radius silica nanoparticle. This particle was initially trapped using the SLD beam and parametrically cooled to sub-kelvin temperature and data on the reheating rates at different pressures were collected. Subsequently, the particle was transferred to the laser beam using the methodology developed earlier \cite{RahmanOptica2020}. Once the particle was successfully transferred to the laser beam, reheating rates under laser levitation were measured and the trapping laser power at the focus was $105~$mW. Solid black lines represent theoretical fits of the form $a_{ph}^q+a_2P_g$.}
    \label{fig3}
\end{figure}

We now present measurements of heating rates on the same particle levitated either by the SLD or the laser as a function of gas pressure. The laser power at the focus was $105~$mW. To make these measurements the particle was transferred between the laser and SLD beam at high pressure before reducing the pressure to make the measurements shown here \cite{RahmanOptica2020}. These experiments were performed on a larger particle (radius $r=70\pm 20~$nm) where the recoil force is larger than that for the $55$ nm particle as the scattering cross section scales as $r^6$. The gas damping will have a smaller effect as it scales with radius as $r^{-1}$. Fig. \ref{fig3}, are plots of the reheating rates of this larger particle for both the SLD and the laser along each axis as a function of the residual gas pressure. Parametric feedback cooling along the $z~$axis was not as efficient as for the 55 nm particle and the lowest CM temperature on this axis was $\approx 1~$K. This meant that the particle was more susceptible to heating from parametric noise along this axis \cite{Jain2016} which can be seen in figure 4. Like the smaller particle, the reheating rates of this larger particle also show a directional dependence which reduces with increasing pressure as heating from gas collisions dominates. For this larger particle, the pressure at which the gas heating rate exceeds the photonic heating rates is now approximately $7\times 10^{-6}~$mBar which is about three times higher than that for the $55$ nm particle. From fits to the SLD data we determine experimental photonic heating rates of $0.12\pm 0.02~$K/s, $0.74\pm 0.07 ~$K/s, $1.03\pm 0.25~$K/s along the $x$, $y$ and $z$ directions. From Eq. (\ref{eqn7}), the theoretical heating rates are determined to be $0.12\pm 0.06~$K/s, $0.44\pm 0.21~$K/s and $1.02\pm 0.60~$K/s along the $x,~y~ \&~ z$ axes, respectively and again the $x$ and $z$ axes are consistent with that experimental values for $x$ and $z$ but not in the $y$. Moreover, these rates are twice than those of $55$ nm particle presented earlier and agree, within the uncertainty determined by the spheres radii. The experimental gas heating rate found from the fit is $0.014\pm0.004~$K/s at $5\times 10^{-8}~$mBar. This is again inline with a separate calculation of the damping rate of particles of these size. As expected the heating rates due to the laser are almost an order of magnitude lower than that due to the SLD. In addition, the laser heating rates are dominated by the gas collisions, even at the lowest pressures, and there is no directional dependence observed and we fit only one line. From the fit we retrieve an offset i.e. the photonic heating rate of $0.02\pm 0.01~$K/s. From theoretical calculations we have $\frac{dE}{dt}|_{ph}^x=0.018\pm 0.018~$K/s, $\frac{dE}{dt}|_{ph}^y=0.036\pm 0.040~$K/s $\frac{dE}{dt}|_{ph}^z=0.124\pm 0.140$K/s. The experimental value is within the uncertainties of theoretical results.

If the particle can be held at lower pressures where gas heating is not significant, and additionally, where other sources of heating do not become important, the particle should come to an equilibrium temperature determined by the thermal light source, in this case, the bulk temperature of the SLD\cite{RahmanPRL2021,Einstein1909}. Damping of the particle motion occurs via the Doppler effect in the scattering of photons from the particle. The Doppler frequencies of the incident and scattered photons are $\omega(1+\frac{\mathbf{v}.\mathbf{\Theta_i}}{c})$ and $\omega(1+\frac{\mathbf{v}.\mathbf{\Theta_s}}{c})$, where $\mathbf{v}$ is the velocity of the particle. For a thermal light source with a bulk temperature of $T$, in the moving frame of the particle, the Doppler effect including the shift in frequency and solid angle can be represented by an effective temperature \cite{HeerKohlPR1968,PeeblesPR1968} $T/(1+\beta_{i,s})$, where $\beta_i=\mathbf{v}.\Theta_i/c$ and $\beta_s=\mathbf{v}.\Theta_s/c$. Explicitly, the damping force can be expressed as \cite{HeerKohlPR1968,PeeblesPR1968,RahmanPRL2021}
\small
\begin{eqnarray}
\nonumber
\mathbf{F}&=&\frac{A_{pn}G}{A_w}\exp{(\frac{\mu_c}{k_BT})}\int_0^\pi\int_0^{2\pi}\int_0^{\theta_{mx}}\int_0^{2\pi}\int_{\omega_c}^{\infty}\Bigg[P_r\sigma_{s}\\
\nonumber
&&\times \frac{\omega^2 }{\Omega_{mx}\pi^2c^2}\Biggl(\frac{ \hbar k \mathbf{\Theta}_{i}}{\exp{\Bigl[\frac{\hbar\omega(1+\beta_i)}{k_BT}\Bigr]}}+\frac{\hbar k \mathbf{\Theta}_{s}}{\exp{\Bigl[\frac{\hbar\omega(1+\beta_s)}{k_BT}\Bigr]}}\Biggr)\Bigg]d\omega d\Omega_{i}d\Omega_s \\
&\approx& \frac{2M\gamma^q_{ph}~\mathbf{v}}{1-\exp[{(\mu_c-\hbar\omega_c)/k_BT}]}.
\label{eqn6}
\end{eqnarray}
\normalsize
At equilibrium the loss and the gain in energy must be equal \cite{RahmanPRL2021} e.g. $ \mathbf{F}.\mathbf{v}=\frac{dE}{dt}|_{ph}^q$, where $v_q^2=k_BT_{ph}^q/M$ and the temperature is given by
\begin{eqnarray}
T_{ph}^q= \frac{T}{2}.
\end{eqnarray}
and the motional temperature of the particle thermalizes to the bulk temperature of the photon source when $\hbar\omega_c < \mu_c$. However, when the chemical potential $\mu_c$ is equal to $\hbar\omega_c$ (see Eqs \ref{eqn7} $\&$ \ref{eqn6}), the source becomes a Bose-Einstein condensate of photons which can still produce thermal photons  \cite{KlaersNat2010}. An equilibration time of $2\pi/\gamma_{ph}\approx 2000~s$ along the $z-~$axis for a $70~$nm radius particle is expected. For a laser it is anticipated to be $\approx 1\times 10^5~$s. To reach equilibrium, however, other sources of heating such as vibrations and parametric heating processes needs to be be minimized. 

In conclusion, we have demonstrated the enhanced recoil heating of an optomechanical object via thermal photons when levitated in vacuum using a superluminescent diode source. This noise dominates over other classical sources of noise such as relative intensity noise and pointing instabilities that typically dominate heating within optical traps. We have measured the heating along all three trap axes for two sizes of levitated nanoparticles. These have been compared with heating by gas and from a laser at the same pressures. Future experiments could seek to measure a thermal photon dominated equilibrium final temperature in a deep trap. Here one can consider ion/Paul traps for nanoparticles which are significantly deeper than the tweezer trap used here and are capable of levitating particles in high vacuum without feedback cooling.


%

\end{document}